\documentclass[12pt]{article}
\usepackage[dvips]{epsfig}
\psfigdriver{dvips}

\renewcommand{\baselinestretch}{1.0} 

\newcommand{\ee}{$(e^+e^-)\; $}
\newcommand{\eeee}{$(e^+e^+e^-e^-)\;$}
\newcommand{\Q}{$Q_0\;$}
\newcommand{\eeg}{$(e^+e^-\gamma)\;$} 

\begin{document}

\begin{center}

{\large\bf The APEX/EPOS Quandary: the Way Out via Low Energy Studies}
\vspace{.10in}

{\large James J. Griffin}

{\it Department of Physics, University of Maryland\\ 
  College Park, Maryland 20742-4111, USA\\
griffin@quark.umd.edu}
\end{center}
\begin{flushleft}
\footnotesize
(This talk was presented at the 8th International Conference on Nuclear
Reaction Mechanisms, in Varenna, Italia, June 9-14, 1977,and will appear
in the Proceedings of the Conference, edited by Dr. Ettore Gadioli. This
manuscript is archived at nucl-th/yymmnn.)
\normalsize

 {\bf Abstract}
\vspace{-0.0cm}

\hspace{.25in}  
A scorecard summary of the various data  of the ``Sharp Lepton
Problem'' is presented. The present situation, in which APEX reports
``...no evidence for sharp pairs...'' even as their data exhibits a
sharp pair excess near 800 keV, is discussed. Two kinds of low energy
experiments utilizing non-heavy ion processes are suggested as
means to break the impasse arising from the ambiguity of the present
heavy ion data.

\vspace{-0.5cm}
\begin{center}
\large{\bf 1. Introduction}
\vspace{-0.5cm}
\end{center}
\normalsize 

The history of the ``Sharp Lepton Problem'' (which is the ``\ee
Puzzle'' of the heavy ion pairs extended to include the sharp electrons
observed in ($\beta^+$ + ATOM) collisons), exemplifies the difficulties
in studying  weak signals of unknown origin.  Here one confronts data
for which the physics we know offers no explanations whatsoever. This
is not typical of one's  customary research experience. Some react
skeptically: if we don't understand it it must be spurious; others
become enthused, and sometimes too uncritical.  Meanwhile the
scientific community in which we exist, like its individual members,
also adopts shifting attitudes.  Thus in the late 1980's pair research
was at the top, and no effort or expense was beyond consideration. Now
in the late 1990's, the community seems to want the whole field of
sharp leptons simply to go away and be forgotten. Perhaps we need
success more than we desire truth. Along with these very human
vacillations, scientific standards seem also to bend. Sceptics are
outcasts when enthusiasm reigns, and seers when the chill sets in. In
the meanwhile working researchers must cling to their standards of
objectivity, openess, and integrity, and strive against the emotional
tides to base their judgements only upon the scientific evidence. In
the end physics is grounded in empirical fact, and in the end the real
physical truth will emerge.

\vspace{-0.1cm}
In this spirit we present here a brief review of the Sharp Lepton data
accumulated so far. We offer also a reinterpretation of the negative
first results reported from the APEX experiment, which was proposed to
settle the issue of the heavy ion sharp pairs once and for all. Sadly,
its results are insufficient to that goal, but have been so 
misstated as to obscure that fact. In the end we emphasize the
lepton and gamma alternatives to heavy ions for studing the Sharp
Lepton Problem, and suggest that such cheap and reproducible
experiments may offer the escape from the quandary presented by the
present ambiguity of the heavy ion data.


\vspace{-0.5cm}
\begin{center}
\large{\bf 2. The Various Sharp Lepton Data }
\vspace{-0.5cm}
\end{center}
\normalsize 

Table I lists the various classes of  data relevant to the ``Sharp
Lepton Problem'', viewed over the template of the Quadronium Scenario.
Included are the sharp ($\Gamma\sim$80keV) positrons, which first lead
to the search for the very sharp ($\Gamma\leq$40keV) \ee pairs.
Independently, very very sharp ($\Gamma\leq$3keV) electrons have been
observed in collisions of beta decay positrons with high-Z U and Th
atoms. Also listed are prospective Delbr\"{u}ck resonances, arising
from creation of the \{\Q,Z\} molecular bound state in ($\gamma$, Z)
processes upon high-Z elements.  (Such bound states were envisaged
already in the very earliest \Q phenomenology\cite{grif/91a,grif/91b}
of the EPOS/I pairs from U + Ta collisions.)  Finally the long standing
10$\sigma$ discrepancy in the lifetime for the 3$\gamma$ decay of
orthopositronium is listed because it, among all of the high precision
quantities of QED\cite{grif/95e} should be especially sensitive to the
bound state poles of the \Q particle, whose existence is the central
hypothesis of the ``\Q Phenomenology''.  If this discrepancy persists
after the calculation of the next order QED correction to the lifetime
has been completed, attention would then turn towards more
non-conventional explanations, such as the existence of the bound \Q
particle.

\vspace{-0.6cm}

\begin{center}
\renewcommand{\arraystretch}{1.3}
\begin{tabular}{|l|l|ccccc|}  \hline 
\multicolumn{7}{|c|}{TABLE I: SCORECARD of SHARP LEPTON DATA}\\
\hline \hline
Years &Collaboration:  &EPOS/I & ORANGE &APEX &EPOS/II & \\ \hline
'83-'86 &H.I.Positrons &YES\cite{schw/83,cowa/85}  &YES\cite{koen/87} 
&- &-&\\ \hline
'86-'96 &H.I.Sharp Pairs &YES &YES &(NO)$\Rightarrow$YES 
&(NO)$\Rightarrow$(??) &\\
&[References] &\cite{cowa/86,cowa/87,sala/90} 
&\cite{berd/88,koen/89,koen/93} &\cite{ahma/95};
\cite{grif/97u,grif/97u2} &\cite{ganz/96};\cite{baum/96} & \\ 
&Repeated? & YES\cite{boke/90} &- &- &- &\\\hline \hline
'86-93 &Sharp Electrons & & & & & \\ 
&(from $\beta^+$+Z): &YES\cite{erb/86} &NO\cite{peck/87} 
&NO\cite{wang/87} &YES\cite{barg/89} &YES\cite{saka/88}\\
&Thin? Repeat?&NO &NO & NO &NO &YES\cite{saka/91,saka/92a,saka/93}\\
\hline \hline
'95&\multicolumn{1}{|l|}{ Delbr\"{u}ck ($\gamma$,Z)}    &  & & & &\\
& \multicolumn{1}{|r|}{$\sim$1.8 MeV:}  &YES; 
& \multicolumn{4}{c|}{(Zilges, et al.\cite{zilg/95})}  \\
& $\Gamma_{e+e-}/\Gamma_\gamma$: & ??? 
&\multicolumn{4}{c|}{ (Key to distinguishing \{\Q,Z\} from nuclear 
IPC)}\\ \hline \hline
& Ps$\rightarrow$ 3$\gamma$ Decay: &\multicolumn{5}{c|}
{\Q pole can explain Long Standing 10$\sigma$ QED discrepancy} \\
\hline
\end{tabular}
\end{center}

\begin{flushleft}
\footnotesize
\vspace{-0.4cm}
Table I. The Various data relevant to the Sharp Lepton Problem are
summarized.  Although the most recent heavy ion experiments (APEX and
EPOS/II) report no positive evidence for sharp pairs, both actually
recorded positive signals, as discussed below, and neither can
definitively exclude the lines reported earlier. The non-heavy ion
data, which is accessible at lower cost and which may be more easily
reproduced by independent experimenters, acquires special interest as
the heavy ion efforts flag.

\end{flushleft}
\normalsize

\vspace{-0.1cm}
Of these data, those from the non-heavy ion processes of lepton and
resonant photon scattering upon high-Z atoms are especially interesting
since they are simple, cheap and repeatable as the  heavy ion studies
are not.  In particular, Sakai\cite{saka/91,saka/93} has
repeatedly\footnote{As noted in Table I, two other
experiments\cite{erb/86,barg/89} preceding those of Sakai, et al. have
reported corroborating sharp lepton evidence, and two
others\cite{peck/87,wang/87} report no such evidence. Only Sakai used
thin targets.} reported very very sharp  electron lines emerging from
the irradiation of thin U and Th targets by positrons from energetic
$\beta^+$ decays, with an estimated\cite{grif/95e} cross section of
$\sim$100 mb. Within the Composite \Q Scenario, Sakai's data can be
understood\cite{grif/94} as arising from a supercomposite molecular
bound state, \{\Q,Z\}, of the \Q atom to the nuclear Coulomb field.
Such states would also appear as Delbr\"{u}ck resonances in photon
scattering from high-Z nuclei, of the type recently observed by Zilges,
et al.\cite{zilg/95}.

\vspace{-0.5cm}
\begin{center}
\large{\bf 3. APEX' Bizarre Self-Contradiction }
\vspace{-0.5cm}
\end{center}
\normalsize 

In their brief report\cite{ahma/95} on their extended effort to settle
the question of sharp pairs from high-Z heavy ion collisions, the APEX
collaboration asserts unconditionally that ``No evidence is found for
sharp peaks in the present data.'' But the data plotted logarithmically
in Figure 2 of their own report exhibit a sharp peak near 800 keV of
precisely the type which APEX was seeking, and which they deny having
found. We here discuss these matters briefly, and point out the
erroneous assumption which may have misled APEX to expect more than was
possible and thereby to overlook a result that was less than hoped
for.

\begin{center}
\epsfig{file=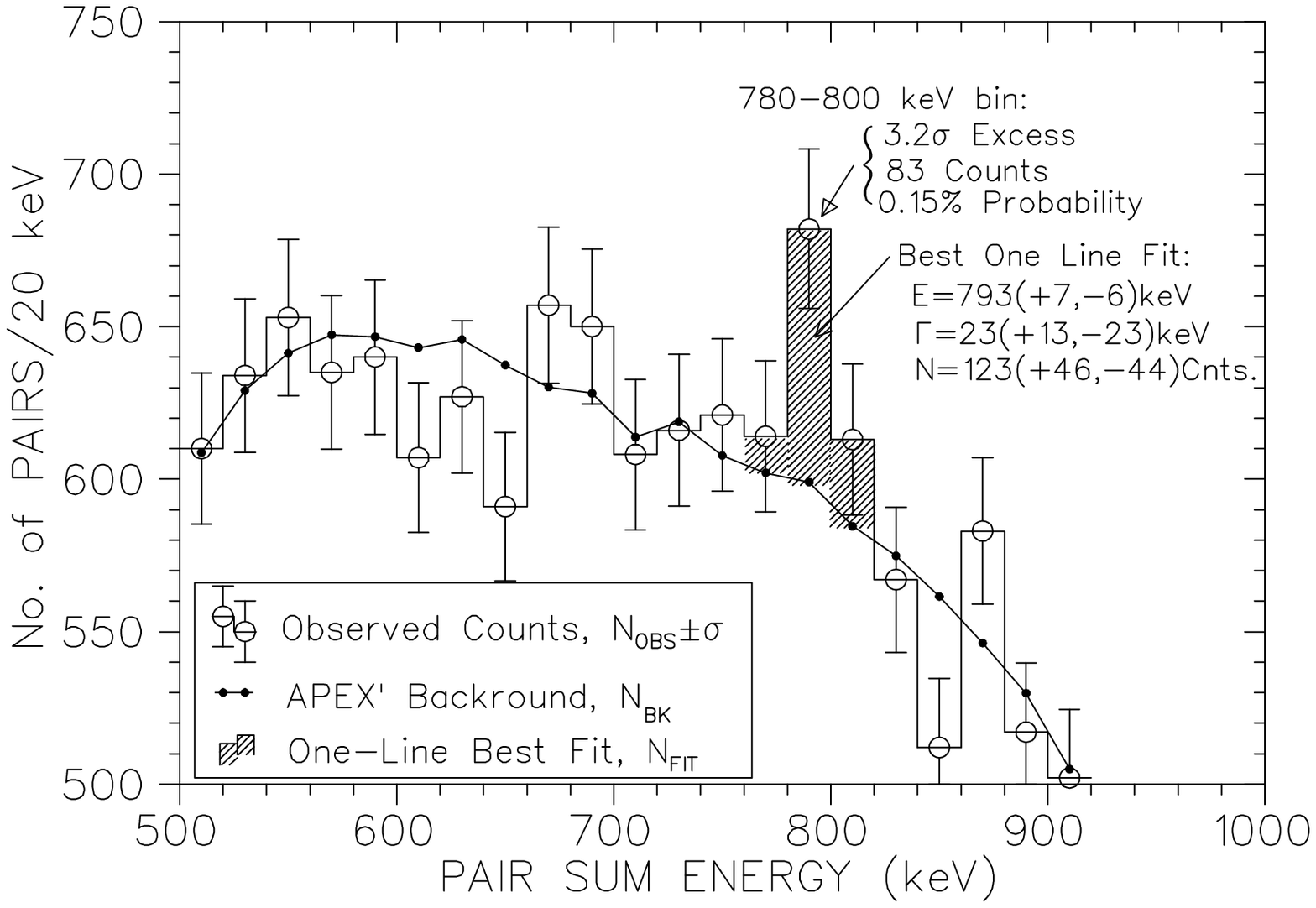,width= 6in,clip=}
\end{center} 

\begin{flushleft}
\footnotesize
\vspace{-0.4cm}
Fig.1. The APEX' data and the APEX' event-event mixed background
published in Fig.2 of Ref.\cite{ahma/95} are plotted. The  
one line best fit (shown shaded), and other statistical analyses
of this data are discussed in the text.
\end{flushleft}  
\normalsize

\vspace{-0.1cm}
The published APEX U +Th data and APEX' event mixed background are
plotted in Fig.1 on a linear scale. A 3.2$\sigma$ excess is
clearly evident in the 780-800 keV bin.  Such an excess is
expected to occur as a fluctuation about once in 700 such single-bin
measurements, or about once in 11 complete 60-bin APEX experiments.

Fig. 1 also shows our 4-parameter  (Background plus One Sharp Line)
best $\chi^2$ fit\cite{grif/97u,grif/97u2} near 800 keV. The best
fitting line has an energy of 793, a width of 23keV, and a strength of
123$\pm$46 sharp pairs. APEX' 1-parameter (Background-Only) fit yields
for the 60 APEX bins a $\chi_{59}^2$ value of  65.76; our 4-parameter
(Background plus One Sharp Line) fit yields $\chi_{56}^2$ = 54.11, a
reduction of 11.65 in $\chi^2$. The probability that the true $\chi^2$
value exceeds these respective values increases from  25\% to 55\% when
the sharp line is allowed,indicating a better quality of fit for the
one-sharp line fit.  Moreover, confidence level analysis of 1-bin,
2-bin and 3-bin groupings all imply that at the 99\% confidence level
there are more than 23 and less than 227 sharp pairs near 790 keV. The
99\% CL lower bounds for groups not including the 790 keV bin are all
negligible (smaller than 2 counts), indicating that at the 99\%
confidence level the APEX data provides evidence for excess sharp pairs
{\it only} in the 790  keV bin.

However one wishes to assess the physical implications of this data, it
is clearly not factually accurate to state, as the APEX report states,
that ``No evidence is found for sharp peaks in the present data''. It
is remarkable that besides APEX' making this assertion which seems to
fly in the face of their own data, they also fail to provide any
statistical analysis whatsoever which supports it.

\vspace{-0.5cm}
\begin{flushleft}
\large{\bf 3.1 APEX and EPOS/I Pair Databases are Comparable }
\vspace{-0.5cm}
\end{flushleft}
\normalsize

Table II compares the APEX and EPOS/I experiments, and their respective
sharp

\vspace{-0.6cm}
\begin{center}
\renewcommand{\arraystretch}{1.2}
\begin{tabular}{|l|c|cc|}  \hline 
\multicolumn{4}{|c|}{TABLE II: COMPARE EPOS/I \& APEX PAIR DATA, 
EFFICIENCIES}\\ \hline \hline
& EPOS/I\cite{sala/90}    &APEX\cite{ahma/95,wola/95}&  \\ \hline
\underline{PAIRS COUNTED} & & &\\ 
TOTAL &- &126K & \\ 
RL(1,n), all n &- &80.1K &\\ 
RL(1,1): $(1e^+,1e^-)$ Only &50K &40.8K&\\
RL(1,1) near 800 keV& 1280 &1480&..per 20 keV\\
Sharp Pairs near 800 keV & 97$\pm$38 &123$\pm$36& \\
Ratio: Sharp/Total RL(1,1) &97/50K  &123/40.8K & \\ \hline
\underline{EFFICIENCIES} & & &(Apex' Proposal)\\ 
positrons: $\epsilon_{e+}$      &10.4\%    &3.7\%  &9.0\%\\
back to back pairs: $\epsilon_{180^\circ}$ &1.4\%     &1.3\%  
&5.6\%\\\hline 
\end{tabular}
\end{center}

\begin{flushleft}
\footnotesize
\vspace{-0.4cm}
Table II. By every quantitative measure, the APEX pair data base is,
for the purpose of confronting EPOS/I's data, at best comparable to
that of EPOS/I, and surely not significantly superior. Therefore, APEX'
weak evidence for a sharp pair line near 800 keV, and its failure 
to reproduce the EPOS/I sharp pair line near 600 keV do not provide
decisive evidence concerning the existence of sharp pairs.
\end{flushleft}
\normalsize

pair counts near 800 keV. It shows that the APEX' 123 sharp pairs
among its 40.8K background pairs of EPOS' RL(1,1) type is roughly
commensurate with the EPOS' count of $\sim$100 sharp pairs among a
total of 50K background pairs: Thus, APEX' $\sim$100 sharp pair count
is roughly what they  should have expected from the EPOS experiment.

But in fact APEX' published expectation (in Fig. 2 of
Ref.\cite{ahma/95}) was much greater: $\sim$2500 sharp pairs near 800
keV. We analyze both experiments in detail in
Ref.\cite{grif/97u,grif/97u2}, and conclude that APEX expectations are
9.3$\times$ too large because of their unsupported, and unsupportable,
assumption that the sharp pair cross section was 5.0$\mu$b/sr and
constant, independent of energy.

\vspace{-0.6cm}
\begin{flushleft}
\large{\bf 3.2 How APEX' Expectations Were Inflated }
\vspace{-0.5cm}
\end{flushleft}
\normalsize 

\vspace{-0.1cm}
Actually, the EPOS/I paper presented\cite{sala/90} definite if
incomplete, evidence for an energy dependent sharp pair production
process, and offered the 5.0$\mu$b/sr value only as an order of
magnitude for  an unspecified ``maximal'' cross section. The APEX'
constant 5.0$\mu$b/sr assumption can therefore not be justified by the
EPOS/I results, or even semantically by the EPOS/I's literal
statements.

\begin{center}
\epsfig{file=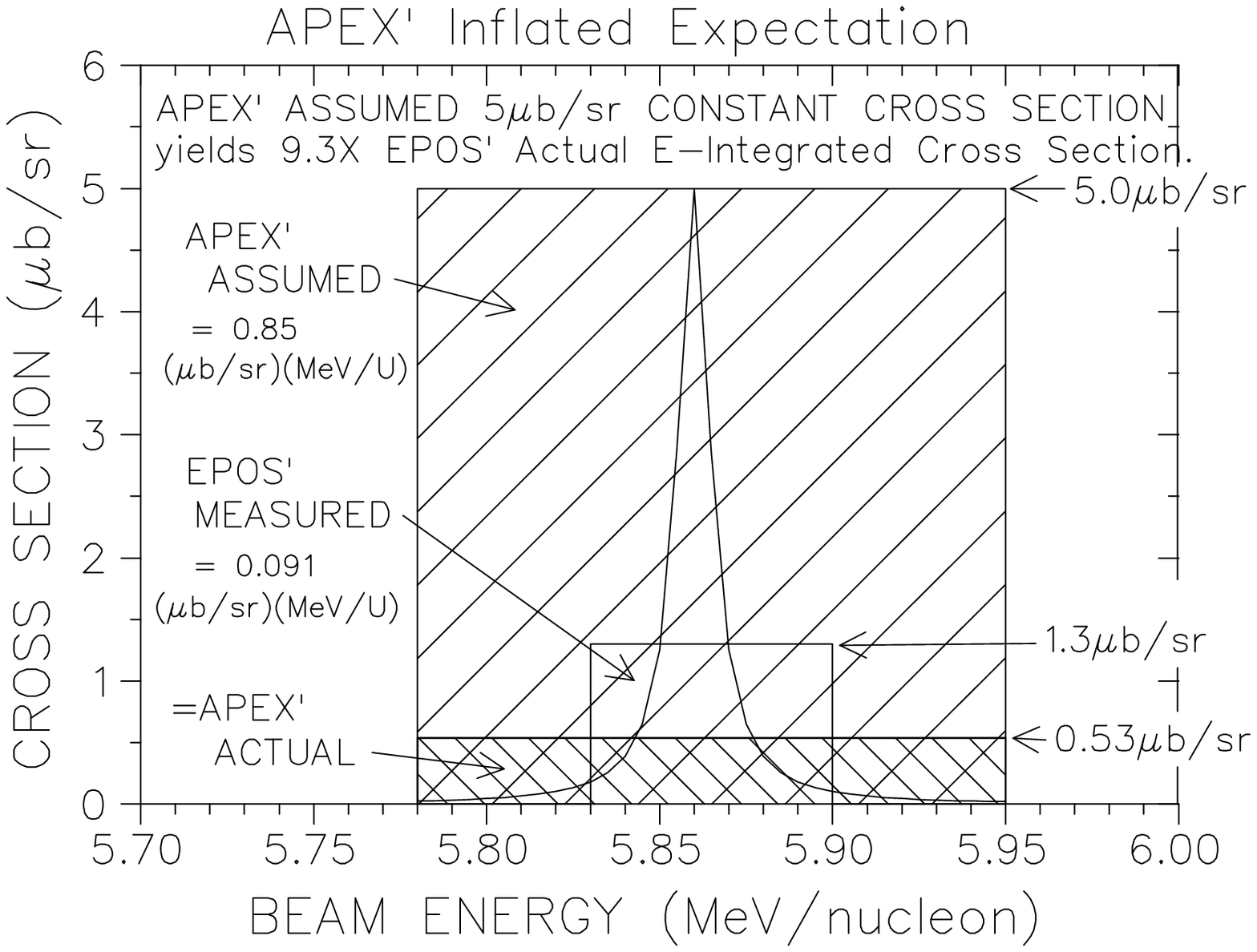,width= 6in,clip=}
\end{center} 

\begin{flushleft}
\footnotesize
Fig. 2 APEX' analysis assumed a constant 5.0$\mu$b/sr sharp pair
production cross section, which yields an energy integrated  cross
section given by the cross hatched area of the figure. This is almost
an order of magnitude larger than the value set by EPOS/I's 
$\sim$100 measured counts.
\end{flushleft}
\normalsize
\newpage

\vspace{-0.1cm}
For a Breit Wigner energy dependence, the EPOS/I sharp pair data
implies no unique value for the sharp pair cross section at all, but
rather a value for the  {\it energy integrated cross section}, which
must be about 0.091($\mu$b/sr)(MeV/nucleon) in order to yield EPOS/I's
$\sim$100 observed sharp pairs. If its maximal value were 5.0$\mu$b/sr,
this cross section would have a width of about 0.02 (MeV/nucleon). Such
a dependence is sketched in Fig. 2. For each different APEX and EPOS/I beam
spread,  the specifically appropriate average pair cross section must
be defined to yield this correct energy integrated value.

\vspace{-0.1cm}
Fig. 2 also exhibits the average cross section of 1.3($\mu$b/sr) (Cf.
Ref.\cite{ganz/96}),  appropriate for the EPOS/I beam energy spread of
0.07(MeV/nucleon), and the value, 0.53($\mu$b/sr),appropriate to APEX'
thicker target beam energy spread of 0.17(MeV/nucleon). This latter
value, 9.3$\times$ smaller than APEX' assumed 5.0$\mu$b/sr, is the
average cross section which EPOS/I's data actually implies for the APEX
experiment. In contrast, APEX' unsupportable 5.0$\mu$b/sr assumption
implies the much larger energy-integrated cross section of 5.0*0.17 =
0.85($\mu$b/sr)(MeV/nucleon), indicated in Fig.2 by the cross hatching,
larger than the actual value by the same factor of 9.3.  Instead of
$\sim$2500 pairs, APEX ought to have been expecting $\sim$270; APEX'
experiment actually counted  123$\pm$46..


\vspace{-0.5cm}
\begin{flushleft}
\large{\bf 3.3 EPOS/II Observed 809 keV Line }
\vspace{-0.5cm}
\end{flushleft}
\normalsize 

Remarkably, the EPOS/II collaboration, which also claims no sharp pair
signal in the only brief report\cite{ganz/96} published so far, reports
elsewhere (in Fig. 6.11 of Ref.\cite{baum/96}) a sharp excess of pairs
at 609 kev, under the same selection conditions as were used by
EPOS/I.  Since this is precisely the energy of a line reported earlier
by EPOS/I\cite{sala/90}, the failure to discuss this observation in
detail in Ref.\cite{ganz/96} is an omission which one hopes will be
rectified in a later publication.

\vspace{-0.5cm}
\begin{center}
\large{\bf 4. Non-Heavy Ion Alternative Studies }
\vspace{-0.5cm}
\end{center}
\normalsize

In Table I, the evidence of Erb et al.\cite{erb/86}, Bargholz, et
al.\cite{barg/89}, and Sakai, et al.\cite{saka/88}, that leptons of
sharp energy emerge from collisions of few MeV positrons with high-Z
atoms opens an experimental window upon the Sharp Lepton Problem which
is alternative to studies with high-Z heavy ion collisions.  Since all
of Sakai's studies have been carried out with positrons whose energy
distribution is set by his energetic $\beta$ emitters, they provide no
evidence as to which positron energies are most effective in their
production. It is therefore a matter of urgency to verify the Sakai
phenomenon with beams of leptons whose energy is well-controlled.

\vspace{-0.5cm}
\begin{flushleft}
\large{\bf 4.1 \Q Spotlights 4-Lepton Box Diagrams in QED }
\vspace{-0.5cm}
\end{flushleft}
\normalsize 

The Quadronium Scenario hangs upon the assumption that the four lepton
\eeee system is strongly (relativistically !) bound.  The resulting
effect upon QED is portrayed in Fig. 3, which shows that if \Q has
bound states, then any QED diagram which contains a 4-lepton ``box''
diagram requires that the corresponding integration over the 4-lepton
continuum must be corrected by the addition of a pole term from each
such bound state, as diagrammed in Fig. 3(c).

\vspace{-0.1cm}
It is obvious from Fig.3 that light upon light scattering will be a
resonant process when the Quantum numbers of the two photons are equal
to those of an eigenstate of \Q. Then  it also follows that
Delbr\"{u}ck scattering (in which two of the photons of Fig 3(b) or Fig
3(c) are replaced by Coulomb interactions with a nuclear Coulomb
field), will also exhibit resonances at incoming photon energies equal
to to an eigenenergy of the \{\Q,Z\} supercomposite bound system, given
by a sharp pair sum energy less the (small\footnote{As the
phenomenolgy\cite{grif/91a} of the U+Ta data indicates it to be.})
\{\Q,Z\} binding energy. If only one (presumably an s-state) state of
\{\Q,Z\} is bound, then resonance will have the spin and parity,
(J$\pi$) of the \Q eigenstate. Then  when the \Q state has (J$\pi$) =
(1$^-$) the excitation of the resonance will be favored; other
multipoles will be excited only with reduced amplitudes.


\begin{center}
\epsfig{file=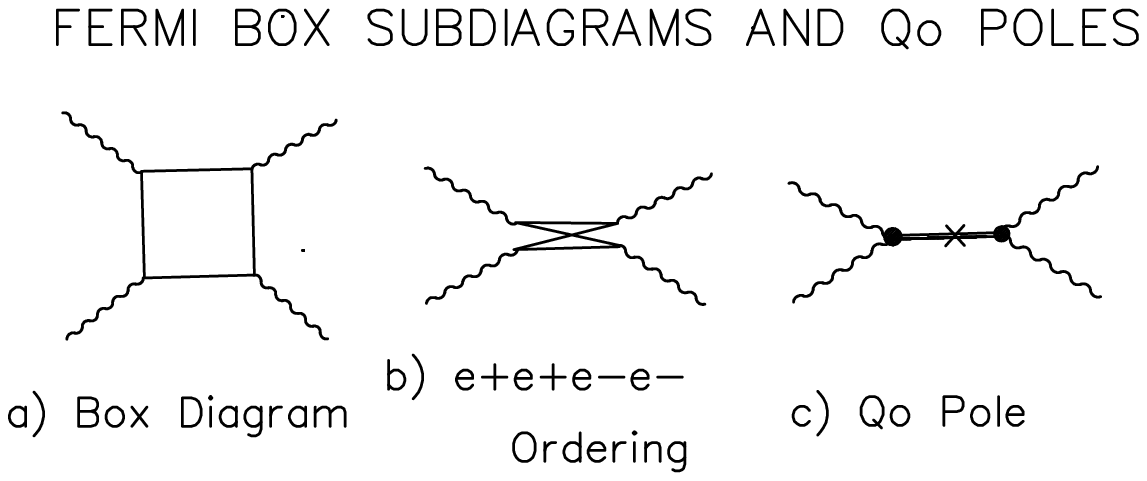,width= 6in,clip=}
\end{center} 

\vspace{-0.6cm}
\footnotesize
Fig.3. (a) The generic four--Fermi box subdiagram of QED; (b) The \eeee
time-ordering of the Fermi box; (c) The \Q--pole, which provides a pole
correction to (b)for each bound state.
\normalsize

\begin{flushleft}
\large{\bf 4.2 Delbr\"{u}ck Scattering Resonances and \Q }
\vspace{-0.5cm}
\end{flushleft}
\normalsize


\vspace{-0.1cm}
Indeed, three resonances have already been observed\cite{zilg/95} near
1.8 MeV in ($\gamma$, U) scattering, and have been interpreted as
conventional nuclear excited states in the U target. But in the
Quandronium Scenario, any one of them may be due to a \{\Q,U\} bound
state rather than a nuclear state. How is one to ascertain the
difference?

\vspace{-0.1cm}
One qualitative distinction is expected to be the branching ratio of
the decay by pair emission as compared to the decay by photon emission.
Nuclear ( 1$^-$) excitations decay to the ground state by emitting a
photon, which if sufficiently energetic may occasionaly produce a pair.
\Q, on the other hand, is most likely to decay to \eeg, yielding an \ee
pair of the total energy when the decay photon is replaced by a Coulomb
interaction with the nuclear charge. Thus one expects pair emission to
be dominant for the \{\Q,U\} bound state, and photon emission for the
nuclear excited state. It is for this reason that the need for
branching ratio evidence is emphasized in the Data Scorecard in Table
I. Additionally, the \{\Q,U\} bound state energy, in constrast with a
nuclear 

\newpage
excitation, should be essentially independent of neutron number
of the nucleus, Z.

\vspace{-0.5cm}
\begin{flushleft}
\large{\bf 4.3 Delbr\"{u}ck \Q Creation and Sakai's Sharp Electrons }
\vspace{-0.5cm}
\end{flushleft}
\normalsize

The incoming photon of the Delbr\"{u}ck scattering can also be
delivered (virtually) to the atom in a bremsstrahlung scattering of a
lepton, as, e.g., in Sakai's positron irradiation. Then the graphs of
Fig.4(c) are the relevant ones. When the \Q is created bound to the
nucleus the compound ssystem,  because of the large mass of the
nucleus, is essentially at rest in the laboratory frame of the target.

\begin{center}
\epsfig{file=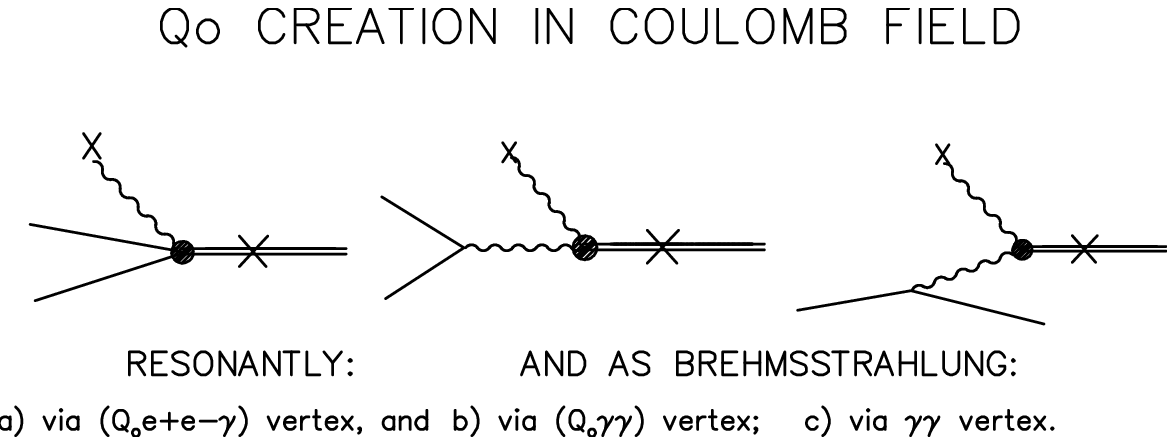,width= 6in,clip=}
\end{center} 

\footnotesize
\begin{flushleft}
Fig.4. \Q creation by leptons can also occur by replacing the incoming
photon of Delbr\"{u}ck scattering process by a bremsstrahlung photon
from a scattered lepton as in Fig.4(c). (In addition the resonant
processes diagrammed in (a) and (b) may occur when an incoming positron
of the correct energy correlates or annihilates, respectively, with one
of the atomic electrons.)
\end{flushleft}
\normalsize

\vspace{-0.1cm}
One might therefore think that this situation is made to order to
explain also  Sakai's very very sharp electrons, whose width requires a
source source stringently at rest (i.e., with a  a kinetic energy for
the \Q source particle of less than 3 eV)  in the lab frame.  But that
is incorrect, because the pair emitted from a bound \{\Q,Z\} state will
exhibit the same effects of the nuclear Coulomb field as have already
been described and observed in the EPOS/I 's U+Ta
data\cite{cowa/87,sala/90,grif/91a,grif/91b}; namely, a $\pm$ shifts of
$\sim$10$^2$ keV in the separate positron/electron energies which
shifts vary with the distance of the decaying \Q from the nucleus. The
latter property is crucial here, because it generates spreads in the
separate lepton energies of the order of 10$^2$ keV, which preclude its
providing Sakai's narrow ($\leq$3 keV) electron lines.

\vspace{-0.1cm}
How then can a scattering process create a \Q particle at rest in the
lab, even when it is stipulated that a resonant bound state of \{\Q,Z\}
is available? One naturally thinks of those events which lead to
slightly unbound \{\Q,Z\} states which can break up and release \Q,
later to decay far from the nucleus.  But these too provide an
insufficient explanation:  in the breakup process, the light \Q
acquires essentially all of the breakup energy as kinetic energy. Then
only states within a few eV (NOT keV!) of the breakup threshhold could
produce such sharp pairs as Sakai observes.

\newpage
\vspace{-0.6cm}
\begin{flushleft}
\large{\bf 4.4 Sakai's  Electrons Stretch the \Q Scenario to Its Limit }
\vspace{-0.5cm}
\end{flushleft}
\normalsize 

\vspace{-0.1cm}
One needs more: a mechanism for the emerging \Q to get rid of its
kinetic energy as it separates from the nucleus, so that it can come to
rest at a point outside of the nuclear Coulomb field. Remarkably, the
\Q Scenario can provide\cite{grif/94} such a process, the ``Viscous
Breakup'' of the slightly unbound \{\Q,Z\} state.  In this process the
\Q, passes through the electron cloud of the U atom, 
=600fm to 7*7ao/92= 3X10+4 fm, where ao 
fm....  $\sim$3X10$^4$ fm.  delivering its kinetic energy into
excitation energy of  the atomic electrons, and emerging from the atom,
(and therefore from the volume where the screened nuclear Coulomb field
is non-zero), with a negligible velocity.

\vspace{-0.1cm}
Such a description requires phenomena which exploit each of the four
distinct length scales of the \Q Scenario: the nuclear ($\sim$10 fm),
\Q (Radius = Compton wavelength, $\sim$10$^2$ fm), Supercomposite Bound
State (Radius, R$_o\sim$ 10$^3$ fm), and the Bohr radii scale
($\sim$10$^4$ fm) of high-Z atoms.  In this way, the explanation of the
Sakai pairs pushes the \Q Scenario perhaps to it very limits.

\vspace{-0.1cm}
Some may view this unfavorably, recoiling against such a ``stretching''
of the hypothesis' possibilities. To  the contrary, we insist that
although two alternative phenomenologies are always preferable to one
phenomenology, one phenomenology is infinitely better than none at
all.  Since we here face this array of Sharp Lepton data which we know
how to summarize under only one phenomenology, the \Q Scenario, we are
obliged to explore all of its possibilities, searching both for a
killer datum which it cannot encompass, and for predictive implications
which can be tested in new experiments.

\vspace{-0.1cm}
For the Sakai sharp lines this procees succeeds wonderfully: The Sakai
lines do not contradict the \Q Scenario, but  instead provide two very
specific verifiable inferences:  (a) that each sharp Sakai electron
accompanies a partner positron which has the same narrow energy
distribution; (b) that (because the diagrams of Fig.4(c) are
indifferent to the charge sign of the scattered lepton) Sakai's sharp
electrons should be found not only in collisions of positrons with
high-Z atoms, but also of electrons upon the same elements, and with
the same cross section as for the electrons.

\vspace{-0.6cm}
\begin{center}
\large{\bf 5. Recommendation: Study  e$^-$'s + (U, or Th) }
\vspace{-0.5cm}
\end{center}
\normalsize 

\vspace{-0.1cm}
The outcome is the prediction that beams of few MeV electrons upon U
and Th atoms should produce sharp {\it positrons} (the decay partners
of Sakai's sharp electrons) of energy 330.1 keV and width $\leq$3 keV,
with a cross section of about 100 mb. Such an experiment will have
large electron backgrounds, analogous to the large positron backgrounds
of Sakai's $\beta^+$ iradiations, but its positrons arise only from
pair production and \Q decay. Since Sakai's positrons eject many
electron from the target atoms, requiring his sharp electrons had to be
observed above a large electron background, the electron beam
experiment promises a smaller positron background to the sharp positron
line being sought than was Sakai's electron background to his sharp
electron lines.

\newpage
\begin{center}
\large{\bf 6. Summary and Conclusions }
\vspace{-0.5cm}
\end{center}
\normalsize 

\vspace{-0.1cm}
A crucial feature of the \Q Scenario is its ability to unify certain
non-heavy ion processes with the ``\ee Puzzle'' posed by the heavy ion
data.  Here it recommends the study of few MeV electron scattering from
high-Z atoms, and of the branching ratios for the decay of Delbr\"{u}ck
resonances in high-Z atoms.

\vspace{-0.1cm}
These non-heavy ion alternatives are essential in view of the present
impasse arising from the failure of the recent (APEX and EPOS/II) heavy
ion experiments to corroborate or to definitively exclude the earlier
(EPOS/I and Orange) reports of sharp pair lines.

\vspace{-0.1cm}
{\bf Acknowledgement} The support of the U.S. Department of Energy
under Grant No.DE-FG02-93ER-40762 is  acknowledged.

\vspace{-0.6cm}
\end{flushleft}

\renewcommand{\baselinestretch}{0.8}
\small
\vspace{-0.1cm}

\begin{thebibliography}{10}

\bibitem{grif/91a}
J.~J. Griffin, {\it Intl.J.Mod.Phys.} {\bf A6},  1985  (1991).

\bibitem{grif/91b}
J.~J. Griffin, {\it Phys.Rev.Lett.} {\bf 66},  1426  (1991).

\bibitem{grif/95e}
J.~J. Griffin, {\it Can.J.Phys.} {\bf 74},  527  (1996), and 
earlier references cited therein.

\bibitem{schw/83}
J. Schweppe {\it et~al.}, {\it Phys.Rev.Lett.} {\bf 51},  2261  (1983).

\bibitem{cowa/85}
T. Cowan {\it et~al.}, {\it Phys.Rev.Lett.} {\bf 54},  1761  (1985).

\bibitem{koen/87}
W. Koenig {\it et~al.}, {\it Z.Phys.} {\bf A328},  129  (1987).

\bibitem{cowa/86}
T.~E. Cowan {\it et~al.}, {\it Phys.Rev.Lett.} {\bf 56},  444  (1986).

\bibitem{cowa/87}
T.~E. Cowan {\it et~al.},  in {\em Physics of Strong Fields}, edited 
by W. Greiner (Plenum Press, New York, 1987), p.\ 111.

\bibitem{sala/90}
P. Salabura {\it et~al.}, {\it Phys.Lett.} {\bf B245},  153  (1990).

\bibitem{berd/88}
E. Berdermann {\it et~al.}, {\it Nucl.Phys.} {\bf A488},  683c  (1988).

\bibitem{koen/89}
W. Koenig {\it et~al.}, {\it Phys.Lett.} {\bf B218},  12  (1989).

\bibitem{koen/93}
I. Koenig {\it et~al.}, {\it Z.Phys.} {\bf A346},  153  (1993).

\bibitem{ahma/95}
I. Ahmad {\it et~al.}, {\it Phys.Rev.Lett.} {\bf 75},  2658  (1995).

\bibitem{grif/97u}
J.~J. Griffin, Univ. of MD. PP No.97-087 (Feb.1997), nucl-th/9703041.

\bibitem{grif/97u2}
J.~J. Griffin, Univ. of MD. PP. No.97-080 (Feb.1997), nucl-th/9703006.

\bibitem{ganz/96}
R. Ganz {\it et~al.}, {\it Phys.Lett.} {\bf B389},  4  (1996).

\bibitem{baum/96}
J. Baumann,   (1996), dissertation (Univ. Heidelberg): Report No.
GSI-96-05.

\bibitem{boke/90}
H. Bokemeyer,   (1990), habilitation thesis (U.of Frankfurt): Report
  No.GSI-90-11.

\bibitem{erb/86}
K. Erb {\it et~al.}, {\it Phys.Lett.} {\bf B181},  52  (1986).

\bibitem{peck/87}
R. Peckhaus {\it et~al.}, {\it Phys.Rev.} {\bf C36},  83  (1987).

\bibitem{wang/87}
T.~F. Wang {\it et~al.}, {\it Phys.Rev.} {\bf C36},  2136  (1987).

\bibitem{barg/89}
C. Bargholz {\it et~al.}, {\it Phys.Rev.} {\bf C40},  1188  (1989).

\bibitem{saka/88}
M. Sakai {\it et~al.}, {\it Phys.Rev.} {\bf C38},  1971  (1988).

\bibitem{saka/91}
M. Sakai {\it et~al.}, {\it Phys.Rev.} {\bf C44},  R944  (1991).

\bibitem{saka/92a}
M. Sakai {\it et~al.},  in {\em Nuclear Physics of Our Times,}, edited 
by A.V.Ramayya (World Scientific, Singapore, 1993), pp.\ 313--321, see 
also U. of Tokyo, I.N.S. Report No.957,Dec. 1992.

\bibitem{saka/93}
M. Sakai {\it et~al.}, {\it Phys.Rev.} {\bf C47},  1595  (1993).

\bibitem{zilg/95}
A. Zilges {\it et~al.}, {\it Phys.Rev.} {\bf C52},  R468  (1995).

\bibitem{grif/94}
J.~J. Griffin,  in {\em Topics in Atomic and Nuclear Collisions}, 
edited by A. Calboreanu and V.Zoran (Plenum Press, New York, 1994), 
p.\ 419.

\bibitem{wola/95}
M.~R. Wolanski, Ph.D. thesis, U. of Chicago, Aug. 1995.

\end{thebibliography}

\end{document}